\documentclass[prd,aps,floats,amssymb]{revtex4}

\usepackage{epsfig}
\renewcommand{\email}[1]{\tt #1}

\begin{document}
\vspace*{1.cm}
\title{Dark Fluid: a complex scalar field to unify dark energy and dark matter\vspace*{1.cm}}

\author{Alexandre Arbey\footnote{E--mail: \email{arbey@obs.univ-lyon1.fr}}\vspace*{0.5cm}}

\affiliation{Centre de Recherche Astronomique de Lyon (CRAL), 9 Avenue Charles Andr\'e, 69561 Saint Genis Laval Cedex, France\vspace*{0.5cm}\\}
\affiliation{\vspace*{0.5cm}Department of Physics, Mount Allison University, 67 York Street, Sackville, New Brunswick, Canada E4L 1E6\vspace*{1.cm}}

\begin{abstract}
\vskip 0.5cm
\noindent In this article, we examine a model which proposes a common explanation for the presence of additional attractive gravitational effects -- generally considered to be due to dark matter -- in galaxies and in clusters, and for the presence of a repulsive effect at cosmological scales -- generally taken as an indication of the presence of dark energy. We therefore consider the behavior of a so-called dark fluid based on a complex scalar field with a conserved $U(1)$-charge and associated to a specific potential, and show that it can at the same time account for dark matter in galaxies and in clusters, and agree with the cosmological observations and constraints on dark energy and dark matter.
\vspace*{2.cm}
\end{abstract}

\vskip 1.cm
\maketitle

\section{Introduction}
\noindent After many years of considering that the Universe is filled with a dark matter made of Weakly Interacting Massive Particles (WIMPs) and with a dark energy causing an acceleration of the expansion of the Universe, we have still no direct evidence of their existences. Moreover, observations of supernov\ae~of type Ia tend to cause trouble to usual dark energy models \cite{SNIa}, and open the way to new kind of models or analyses to explain the observed repulsing effects \cite{phantom,coley}.\\
Alternative models exist to explain the cosmological observations, and in particular some of them try to solve the dark energy and dark matter problems by unifying both components into a single ``dark fluid''. We can for example note that the Generalized Chaplygin Gas model \cite{chaplygin} follows this idea and is presently under scrutiny. We have shown in \cite{arbey_darkfluid} that building such a dark fluid model is very difficult, as the model has to be in agreement with many observational constraints, and especially has to explain at the same time the cosmological repulsive effects and the local binding gravitational effects in the recent Universe. As scalar field-based models for dark energy and dark matter exist in the literature \cite{quintessence,quintessence2,hessence,matter_field,fuzzy}, it seems interesting to study unifying dark fluid models based on scalar fields. This idea has been proposed in \cite{padmanabhan}. The crucial question however concerns the form of the potential of the scalar field.\\
In this article, we will consider a complex scalar field and propose an adequate form for its potential. We will show that this scalar field can potentially explain correctly the observations of galaxy rotation curves of spiral galaxies and the presence of strong binding gravitational effects in clusters. We will also consider the cosmological behavior of the dark fluid and show that it agrees with the cosmological constraints and observations. We will finally conclude by suggesting some further directions of investigation beyond this study.

\section{Description of the model}
\noindent We consider here a scenario in which the dark fluid can be accounted by a complex scalar field associated to a conserved charge. We focus on the idea that the scalar field evolves with a quasi-homogeneous density in the Early Universe, then produces cluster and galactic halos through Bose condensation and provides repulsing effects outside the Bose condensates. The Lagrangian density of the scalar field reads
\begin{equation}
\mathcal{L} = g^{\mu\nu} \partial_\mu \phi^\dagger \partial_\nu\phi - V(\phi) \;\;,
\end{equation}
and we assume that its potential $V$ is invariant under the global symmetry 
\begin{equation}
\Phi \rightarrow \Phi'=e^{i \theta} \Phi \;\;.
\end{equation}
The principal issue to elaborate a dark fluid model is the choice of a potential. Considering \cite{arbey_darkfluid}, we can conceive that a model for a dark fluid would be promissing if it has a negligible density at the Big-Bang Nucleosynthesis (BBN) time, if it provides a matter behavior at the time of recombination and of structure formation, and if today its behavior is cosmologically repulsive, and attractive in clusters and galaxies. This kind of behavior seems however very difficult to achieve.\\
We can nevertheless consider the studies of complex scalar fields with quadratic and quartic potentials which were performed in \cite{arbey_quadratic,arbey_cosmo,arbey_quartic}. In those articles, we showed that these potentials leads to a matter behavior from the time of the recombination until today, and are in agreement with the constraints of BBN \cite{BBN}. Hence, we can consider a potential containing a quadratic term, which will provide a matter behavior. However, this potential has to be modified to give a repulsive effect today. Several quintessence potentials, such as a decreasing exponential \cite{exponential} or an inverse square law \cite{inversesquarelaw}, could provide an adequate behavior. In particular, the decreasing exponential seems adequate, and the potential reads in this case
\begin{equation}
V(\phi) =m^2 \phi^\dagger \phi + \alpha \exp(-\beta |\phi|) \;\;.
\end{equation}
$\alpha$ and $\beta$ can then be chosen, so that the quadratic term is dominant until the late time of the structure formation, and the exponential term becomes non-negligible only later, once $\phi$ has become small. In that way, it seems possible to have a correct cosmological behavior for the fluid. In galaxies, the densities are quite large, and so is $\phi$, consequently the exponential term can again be negligible. In the interstellar medium, where densities are small, $\phi$ is also small and the exponential term can dominate in the potential and provide a repulsive behavior.\\
This potential could then seem interesting to study, but problems could appear when trying to explain the attractive gravitational effects at the same time in galaxies and in clusters, due to the different scales involved. Another question would concern the way to constraint the parameter $\beta$, which is, apart from the fact that it has to be small, quite unclear. Another choice of potential may then be more appropriate.\\
\\
Thus, we will rather consider the following potential, constituted of a quadratic term and a Gaussian term:
\begin{equation}
V(\phi) =m^2 \phi^\dagger \phi + \alpha \exp(-\beta \phi^\dagger \phi) \;\;.
\end{equation}
This potential can be motivated in the scope of the non-perturbative renormalization group \cite{renorm_group}. The interesting feature of this potential is that the second term has the same kind of behavior than the usual decreasing exponential potential, but we now have a way to determine $\beta$. In fact, when $\phi$ is small, for example outside galaxies, this potential can be approximated by a quartic potential, which is known to lead to a matter behavior \cite{arbey_quartic}. Therefore, a way to constraint $\beta$ would consist in choosing it so that the approximate quartic term leads to a matter behavior at cluster scale, whereas the quadratic term leads to a matter behavior at galaxy scale.\\
\\
Let us now review the constraints on the three parameters $m$, $\alpha$ and $\beta$. We will use for this study the results of \cite{arbey_darkfluid,arbey_quadratic,arbey_quartic}.\\
\\
\begin{itemize}
\item \underline{Galactic scale: determination of $m$}\\
\\
First, we want the mass parameter $m$ to give a matter behavior in galaxies. The Compton wavelength associated to the quadratic term is
\begin{equation}
l_{\mbox{compton}}=\frac{\hbar}{mc}\;\;.
\end{equation}
At local scales, we assume that the complex scalar field has a conserved $U(1)$-charge (see Section~\ref{section_galactic} for more details) and takes the form
\begin{equation}
\phi(\vec{x},t) = \frac{\sigma(\vec{x})}{\sqrt{2}} \exp(-i\omega t) \;\;,
\end{equation}
where $\sigma$ is the norm of the scalar field, $\omega$ is a constant phase, $t$ is the time and $\vec{x}$ the position. With a quadratic potential, it was shown in \cite{arbey_quadratic} that the size of the bosonic halo is then given by
\begin{equation}
l \sim \sqrt{\frac{1}{G^{1/2} \sigma_0}}\, l_{\mbox{compton}}  = \sqrt{\frac{1}{G^{1/2} \sigma_0}} \frac{\hbar}{mc} \;\;,
\end{equation}
where $\sigma_0$ is the value of $\sigma$ at the galactic center. The typical orbital velocity in such a halo is
\begin{equation}
\frac{v}{c}\sim \sqrt{G^{1/2} \sigma_0}\;\;.
\end{equation}
For a typical velocity $v$ of the order of 100 km/s and a typical halo size $l$ of 10 kpc, we can estimate the mass to be about $m \sim 10^{-23}$ eV, which is in agreement with the results of \cite{arbey_quadratic}.\\
\\
\item \underline{Cosmological scales: determination of $\alpha$}\\
\\
We want that today the exponential term of the potential dominates, so that a repulsive behavior could dominate at cosmological scales. From the study of the dark fluid parameters performed in \cite{arbey_darkfluid}, it is convenient to choose
\begin{equation}
\alpha = \rho_{\mbox{dark energy}}^0 = 0.71 \times \rho_0^c \sim 7 \times 10^{-27} \mbox{ kg m$^{-3}$}\;\;,
\end{equation}
where $\rho_0^c$ is the cosmological critical density, so that the exponential term plays the role of dark energy.\\
\\
\item \underline{Cluster scale: determination of $\beta$}\\
\\
Finally, we would like that the exponential term also gives a matter behavior in clusters, {\it i.e.} acts as a quartic potential when the field has a small norm. In this case, the potential reads, at fourth order in $\phi$:
\begin{eqnarray}
V(\phi) &\approx& m^2 \phi^\dagger \phi + \alpha - \alpha \beta \phi^\dagger \phi + \frac{1}{2}\alpha \beta^2 (\phi^\dagger \phi)^2\\
&\approx& \alpha + (m^2-m'^2) \phi^\dagger \phi + \lambda (\phi^\dagger \phi)^2 \;\;,
\end{eqnarray}
with $m'^2 = \alpha \beta$ and $\lambda = \alpha \beta^2 / 2$. We can assume here that $m > m'$. We have shown in \cite{arbey_quartic} that the scalar field condenses on a distance $L$, which we would like to be of the typical size of a cluster ($L$ $\approx$ 1 Mpc). In this case
\begin{equation}
\lambda = \frac{8 \pi G m_{\mbox{eff}}^4 L^2}{c^2}\;\;,
\end{equation}
where $m_{\mbox{eff}}^2=m^2-m'^2$. The typical cluster scale of 1 Mpc then imposes
\begin{equation}
\lambda \sim 10^{-89} \;\;,
\end{equation}
and
\begin{equation}
\beta = \sqrt{\frac{2 \lambda}{\alpha}} \sim 10^{-39} \mbox{ eV$^{-2}$} \;\;.
\end{equation}
We can then calculate $m'$ to confirm that $m > m'$, and show that the effective mass is not very different from the quadratic mass ($m_{\mbox{eff}} \approx 0.999 \,m > 0$).\\
\\
\end{itemize}
Thus, with these choices of parameters, the potential has a single non-degenerate minimum for $\phi = 0$, which eludes the question of global cosmic strings.\\
Considering that the different parameters of the model have been fixed, we can now analyze the behavior of the field at galactic, cluster and finally cosmological scales for these parameters.

\section{Galactic scale}\label{section_galactic}
\begin{figure}[!ht]
\centering
\epsfig{file=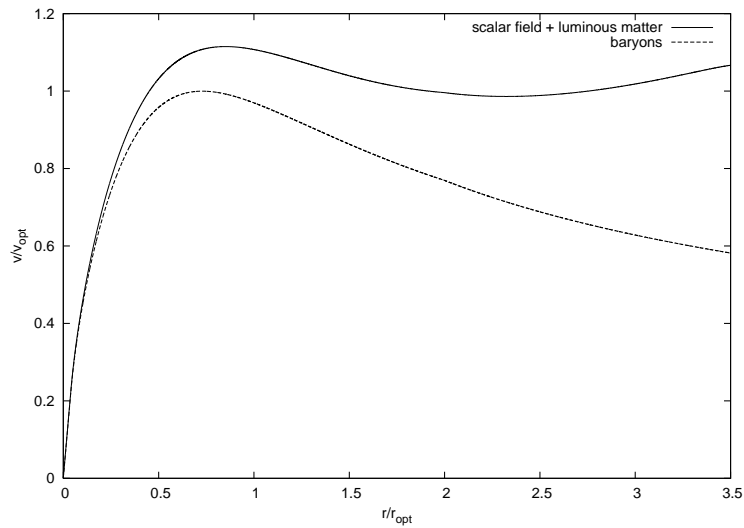,width=7cm,angle=270}
\caption{Rotation curve of a generic spiral galaxy, in presence of a complex scalar field with a uniformly rotation phase accounting for dark matter. The solid line corresponds to an example of rotation curve provided by the complex scalar field plus the baryons, whereas the dashed line shows the contribution of the baryons only.}
\label{rotation_curve}
\end{figure}

\noindent We will first show that this scalar field can play the role of dark matter at galactic scale. Let us consider that a galaxy evolves in a space represented by a static and isotropic metric
\begin{equation}
d\tau^{2} = e^{2 u} dt^{2} -e^{2 v} \left(dr^{2} + r^{2} d \theta^{2} + r^{2} \sin^{2} \theta d \phi^{2} \right) \;\;,
\end{equation}
where $u$ and $v$ only depend on the position. We will assume a spherical symmetry for the system, as it was already shown that this approximation leads to correct and significant results \cite{arbey_quartic}, and study the rotation curves generated by the model for spiral galaxies.\\
\\
The scalar field is as usual to be considered to have an internal rotation, so that we can write, in a spherical symmetry,
$\phi(\vec{x},t) \propto \sigma(r,t) e^{-i\theta(r,t)}$, where $\theta$ is the internal rotation angle. However, a deeper study of the cosmological behavior of $\theta$ (see in particular equation~(\ref{KGcosmo2}), in which $\omega = \dot{\theta}$) could reveal that when the scalar field behaves like cosmological matter ({\it i.e.} when $\sigma^2 \propto a^{-3}$, with $a$ the cosmological scale factor), the first derivative of $\theta$ with respect to time ($\omega$) is nearly a constant. We thus assume that $\phi(\vec{x},t) \sim \sigma(r,t) e^{-i\omega(r) t}$. As we study a static model, we should also consider that the modulus of the scalar field is time-independent. Besides, we will first assume that $\omega(r)$ is space-independent (but galaxy-dependant). We will discuss this approximation at the end of this section.\\
Thus, following these assumptions, the scalar field has a uniformly rotating phase $\omega$, which corresponds to the simplest possible realization for a scalar field. One could refer to \cite{fuzzy} for an alternative analysis of the behavior of a similar scalar field in galaxies. We can write
\begin{equation}
\phi(\vec{x},t) = \frac{\sigma(r)}{\sqrt{2}} \exp(-i\omega t) \;\;.
\end{equation}
In the case where the internal rotation period $2\pi/\omega$ is much smaller than the rotation period of the galaxy, the Klein-Gordon and Einstein equations read
\begin{equation}
e^{- 2 v} \, \left\{ \sigma'' + \left( u' + v' + \frac{2}{r} \right) \sigma' \right\} c^2  + 
\omega^{2} e^{- 2 u} \sigma - V'(\sigma) = 0\;\;,
\end{equation}
\begin{equation}
2 v'' + v'^{2} + \frac{4 v'}{r} = - \frac{8 \pi G}{c^2} e^{2 v} \left\{ \rho_b + e^{- 2 u} \frac{\omega^{2} \sigma^{2}}{2} + e^{- 2 v} \frac{\sigma'^{2}}{2} + V \right\}
\end{equation}
and
\begin{equation}
\label{galactic_einstein2}
u'' + v'' + u'^{2} + \frac{1}{r} \left( u' + v' \right)  = \frac{8 \pi G}{c^2} e^{2 v} \left\{ 
e^{- 2 u} \frac{\omega^{2} \sigma^{2}}{2} - e^{- 2 v} \frac{\sigma'^{2}}{2} - V \right\} \;\;,
\end{equation}
where
\begin{eqnarray}
V(\sigma) &=& \frac{1}{2} m^2 \sigma^2 + \alpha \exp\left(-\frac{1}{2}\beta\sigma^2\right)\\
V'(\sigma) &=& m^2 \sigma - \alpha \beta \sigma \exp\left(-\frac{1}{2}\beta\sigma^2\right) \;\;.
\end{eqnarray}
We can recover the Newtonian limit by posing $u=\Phi$ and $v=-\Psi$, where $\Phi$ is the usual gravitational potential. A careful examination of $\sigma'$ \cite{arbey_quartic} can reveal that $\sigma'$ can be safely neglected in comparison to $\omega \sigma$.\\
In this limit, the scalar field contributes to the mass of the galaxy by adding an effective density \cite{arbey_cosmo}
\begin{equation}
\rho_{\mbox{eff}}= 4 \dot{\phi}^\dagger\dot{\phi} - 2 V(\phi) = 2 \omega^2 \sigma^2 - m^2 \sigma^2 - 2 \alpha \exp\left(-\frac{1}{2}\beta \sigma^2\right) 
\label{effective_density}
\end{equation}
to the baryon density $\rho_b$. The effective pressure of the scalar field reads
\begin{equation}
P_{\mbox{eff}}= g^{00} \dot{\phi}^\dagger\dot{\phi} - V(\phi) = \frac{1}{2} \omega^2 \sigma^2 - \frac{1}{2} m^2 \sigma^2 - \alpha \exp\left(-\frac{1}{2}\beta \sigma^2\right) \;\;.
\end{equation}
In the Newtonian limit, $\Phi = \Psi$, so that equation (\ref{galactic_einstein2}) leads to
\begin{equation}
\omega^2 \sigma^2 \approx m^2 \sigma^2 + 2 \alpha \exp\left(-\frac{1}{2}\beta \sigma^2\right) \;\;.
\end{equation}
Thus, the effective pressure vanishes, and the scalar field behaves like pressureless matter.\\
\\
In spiral galaxies, the typical density at large radii is of the order of $10^{-21}\mbox{ kg m}^{-3}$, which corresponds to $\sigma \sim 10^{-2} \mbox{ kg$^{1/2}$ m$^{-3/2}$ s}$. Therefore, in this case, $\alpha \exp\left(-\frac{1}{2}\beta \sigma^2\right) \sim 10^{-3000} \mbox{ kg m}^{-3}$ is completely negligible, and only the quadratic term remains. Thus, we are in the same case as the analysis of \cite{arbey_quadratic}, which considered a simple quadratic potential. We have then
$\omega^2 \approx m^2$ and consequently
\begin{equation}
\rho_{\mbox{eff}} \approx m^2 \sigma^2 \;\;.
\end{equation}
For $m = 10^{-23}$ GeV, we can calculate the internal rotation period, and show that it is of the order of 30 years, which confirms the hypothesis that the internal rotation is much faster than the galactic rotation. Concerning the baryons, should the disk be alone, the rotation curve would be \cite{salucci}
\begin{equation}
v^2_{b}(r)=v^2_b(r_{opt}) \frac{1.97(r/r_{opt})^{1.22}}{\{(r/r_{opt})^2+0.78^2\}^{1.43}} \;\;,
\end{equation}
leading to a baryonic density
\begin{equation}
\rho_b(r)=\frac{1}{4\pi G} \frac{v_d^2(r_{opt})}{r_{opt}^2} \left(\frac{4.38 (r/r_{opt})^{-0.78}}{\{(r/r_{opt})^2+0.78^2\}^{1.43}} - \frac{5.64 (r/r_{opt})^{1.22}}{\{(r/r_{opt})^2+0.78^2\}^{2.43}} \right) \;\;,
\end{equation}
$r_{opt}$ being the optical radius, defined as the radius of the sphere encompassing 83\% of the luminous matter.\\
\\
When solving the equations of motion, we can obtain the result that, for each set of parameters, the system can have a discrete number of possible configurations, corresponding to fundamental and excited state of the Bose condensate \cite{arbey_quadratic}. We can derive the rotation curves from the Newtonian gravitational potential.\\
Restricting us to the fundamental state, it is possible to generate approximately flat rotation curves up to large radii, as shown on figure \ref{rotation_curve}. As the potential in galaxies restricts to the quadratic term, we can refer to \cite{arbey_quadratic} for a more complete study of such a potential at galactic scales.\\
The size and the density of the halos are fixed by the value of $\omega$. Then, to each galaxy corresponds a specific $\omega$. As the Newtonian approximation leads to $\omega^2 \approx m^2$, $\omega$ can only slightly vary, which justify the previous approximation that $\omega$ is space-independent inside a galaxy. However, if we want to study the variation of this $\omega$ between galaxies, we have to consider an extra equation coming from the imaginary part of the Klein-Gordon equation:
\begin{equation}
\omega'' \sigma r + 2 \omega' \sigma' r + \left\{2 + r (u' + v')\right\} \omega' \sigma =0 \;\;.
\end{equation}
We can nevertheless note that, because of the very small variations of $\omega$, this equation has no effect inside a specific galaxy, but can however explain the presence of different effective values of $\omega$ inside a galaxy cluster.\\
\\
To conclude this section, this scalar field may account for dark matter in galaxies, what is confirmed in \cite{fuzzy}. Let us now consider the behavior of the scalar field at cluster scale.

\section{Cluster scale}
\noindent In clusters, one can expect the same kind of behavior as in galaxies, an important difference being that the average densities are smaller. We can consider that the typical density in clusters is around $10^{-25}\mbox{ kg m}^{-3}$. From equation (\ref{effective_density}), we can show that in the cluster case, $\alpha \exp\left(-\frac{1}{2}\beta \sigma^2\right) \sim \alpha$. This term cannot be neglected anymore, but can be approximated by
$\alpha - \frac{1}{2} \alpha \beta \sigma^2 + \frac{1}{8} \alpha \beta^2 \sigma^4$. The potential then reads:
\begin{equation}
V(\sigma) \approx \alpha + \frac{1}{2}(m^2-m'^2) \sigma^2 + \frac{1}{4}\lambda \sigma^4 = \alpha + \frac{1}{2} m_{\mbox{eff}}^2\, \sigma^2 + \frac{1}{4}\lambda \sigma^4 \;\;,
\end{equation}
where $m'^2 = \alpha \beta$, $m_{\mbox{eff}}^2 = m^2-m'^2$ and $\lambda = \alpha \beta^2/2$. In this case, the Klein-Gordon equation becomes
\begin{equation}
\lambda \sigma^2 = (1-2 \Phi)\omega^2 - m_{\mbox{eff}}^2 \;\;.
\end{equation}
This result is similar to the one derived in \cite{arbey_quartic}. We can remark that the Klein-Gordon equation does not depend on $\alpha$. The effective density of the scalar field rewrites
\begin{equation}
\rho_{\mbox{eff}}= 2 \omega^2 \sigma^2 - m_{\mbox{eff}}^2 \sigma^2 - \frac{1}{4}\lambda \sigma^4 - \alpha \;\;.
\end{equation}
The scalar field is then in a Bose condensate state, on the boundaries of which the gravitational potential is
\begin{equation}
\Phi_0 = \frac{1}{2}\left(1 - \frac{m^2}{\omega^2} \right) \;\;.
\end{equation}
As the potential $\Phi_0$ is small, we have $\omega^2 \approx m_{\mbox{eff}}^2$, and the effective density becomes
\begin{equation}
\rho_{\mbox{eff}}= -2 \alpha + \frac{2 m^4}{\lambda} (\Phi_0-\Phi) \mathcal{H}(\Phi_0-\Phi) \;\;.
\end{equation}
The Klein-Gordon and Einstein equations can then be combined into the following Poisson equation \cite{arbey_quartic}
\begin{equation}
\Delta \Phi = 4 \pi G (\rho_b + \rho_{\mbox{eff}}) \;\;.
\end{equation}
Hence, the scalar field has a matter behavior up to the end of the Bose condensate. The value of $\lambda$ has to be chosen so that the size of the condensate is approximately the size of a typical cluster, in the way presented in the introduction. In the case where the luminous matter density can be neglected in comparison to the density of the scalar field -- which is a decent hypothesis at cluster scale -- and if the halo can be considered spherical, the Poison equation reduces to the Lane-Emden equation for a polytrop n=1. We can therefore determine analytically the effective density of the scalar field:
\begin{equation}
\rho_\phi = -2 \alpha + \rho_\phi^{0} \frac{\sin z}{z} \;\;,
\end{equation}
where $ \rho_\phi^{0}$ is the scalar field density at the center of the cluster, and $z = r / L$, $L$ being the typical size of a cluster. The effective size of the bosonic halo is then:
\begin{equation}
R = \pi \left\{ {\displaystyle \frac{\lambda}{8 \, \pi \, G}} \right\}^{1/2} \, {\displaystyle \frac{1}{m_{\mbox{eff}}^{2}}}\;\;,
\end{equation}
and the total mass of the cluster is determined by the value of $\sigma$ at the center.\\
Knowing that $\rho_\phi^{0} \approx 10^{-25}\mbox{ kg m}^{-3}$ and $\alpha \approx 7 \times 10^{-27}\mbox{ kg m}^{-3}$, the exponential term will have only a slight effect inside the Bose condensate, but far away from the center of clusters, beyond the end of the Bose condensate, only this term remains in the effective density:
\begin{equation}
\rho_{\mbox{eff}} \approx -2 \alpha \;\;.
\end{equation}
Consequently, outside clusters, the behavior of the dark fluid is repulsive, and so completely similar to that of a cosmological constant.\\
\\
In summary, we can interpret our results in this way: where baryons are gathered in sufficient densities, the scalar field is excited and condenses into attractive halos, and therefore has a matter behavior, whereas for a low baryonic density, the field is not excited anymore and has a repulsive behavior. We can thus guess that the global behavior of the field is repulsive, in particular at cosmological scale. We will verify this idea in the next section.

\section{Cosmological behavior}
\begin{figure}[!ht]
\centering
\epsfig{file=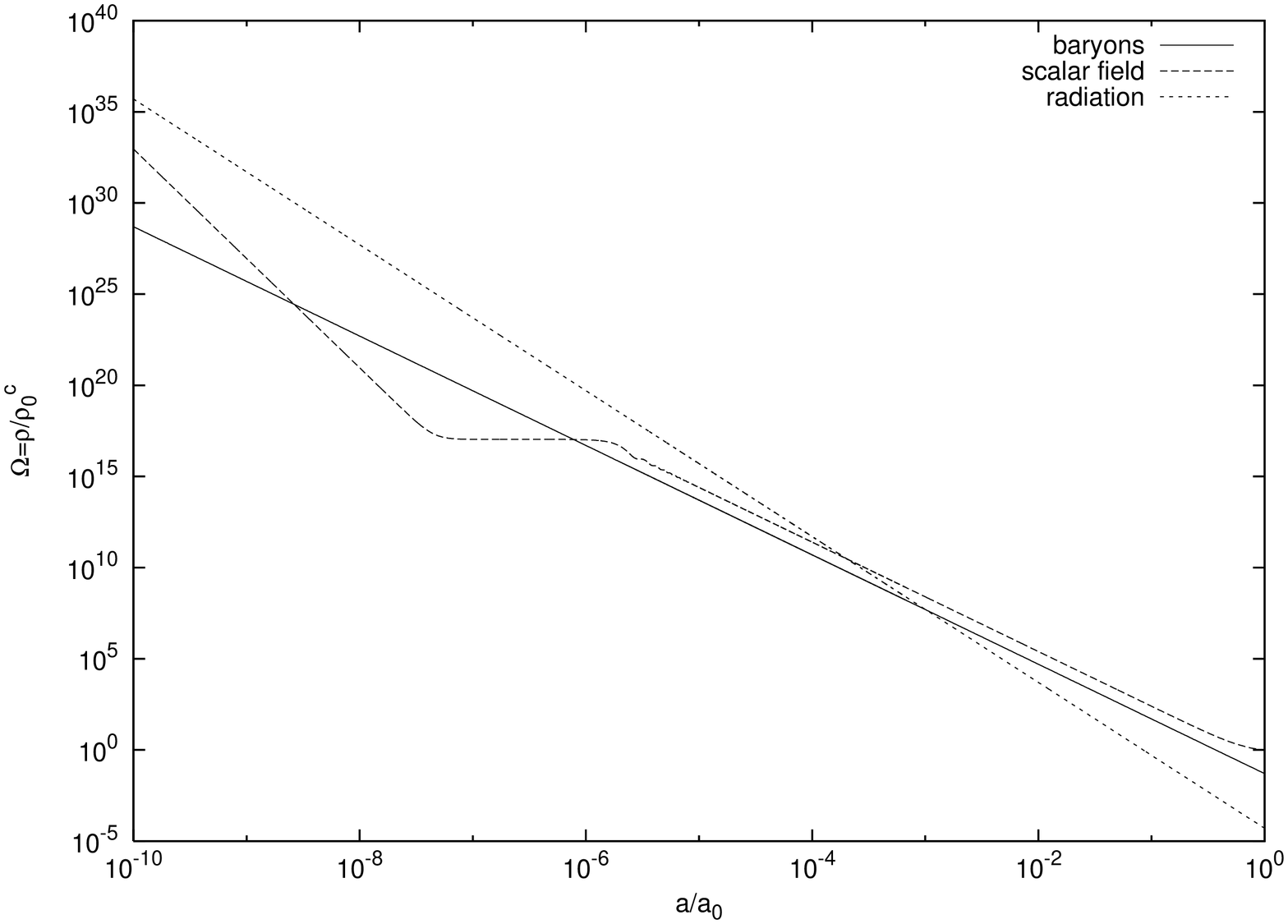,width=6cm,angle=270}\epsfig{file=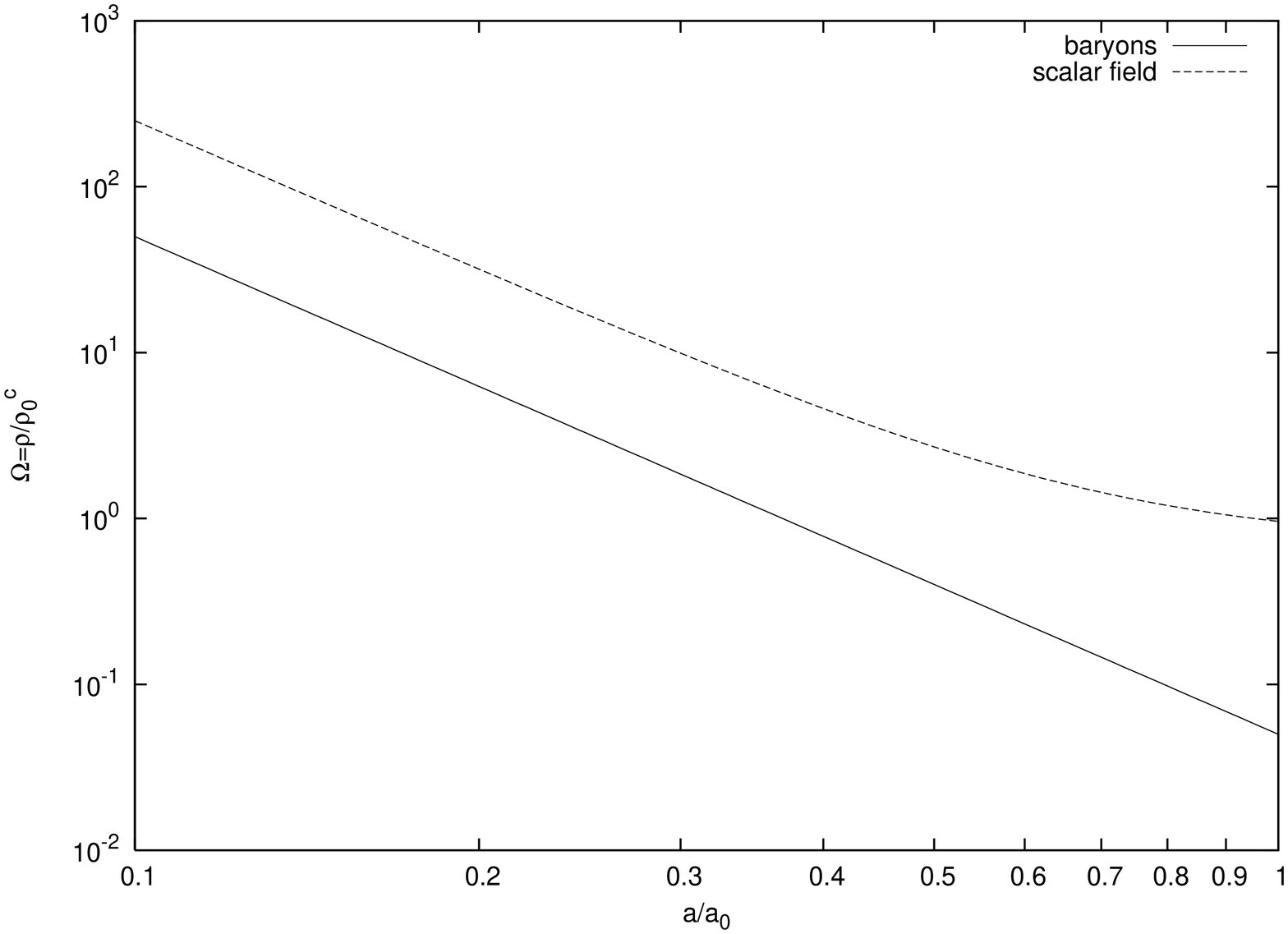,width=6cm,angle=270}
\caption{Cosmological evolution of the density of the scalar field dark fluid in comparison to the densities of baryonic matter and radiation.}
\label{cosmo_evolution}
\end{figure}

\noindent Let us now consider the cosmological behavior of the scalar field, in a flat, homogeneous and isotropic Universe, filled only with radiation, baryonic matter and the dark fluid scalar field. The pressure and density of the scalar fluid can be written
\begin{equation}
\rho_{\phi} = \dot{\phi}^{\dagger} \dot{\phi} + V \left( \phi \right) \;\;,
\end{equation}
and 
\begin{equation}
P_{\phi} = \dot{\phi}^{\dagger} \dot{\phi} -V \left( \phi \right) \;\;,
\end{equation}
where the scalar field has an internal rotation
\begin{equation}
\phi(t) = \frac{\sigma(t)}{\sqrt{2}} e^{i \theta(t)} \;\;.
\end{equation}
The Klein-Gordon and Friedman equations read
\begin{equation}
{\label{KGcosmo1}\displaystyle \frac{d^2 \sigma}{d t^2}} + \frac{3}{a} {\displaystyle \frac{da}{dt}}{\displaystyle \frac{d \sigma}{dt}}
+ m^2 \sigma - \alpha \beta \sigma \exp\left(-\frac{1}{2}\beta \sigma^2\right) - \omega^2 \sigma = 0 \;\; ,
\end{equation}
\begin{equation}
{\label{KGcosmo2}\displaystyle \frac{d \omega}{dt}} \sigma + \frac{3}{a} {\displaystyle \frac{da}{dt}} \omega \sigma + 2 \omega {\displaystyle \frac{d \sigma}{dt}} = 0 \;\; ,
\end{equation}
\begin{equation}
3 H^2 = 3 \left( {\displaystyle \frac{da}{a \, dt}} \right) ^2 =
8 \pi G \left[\rho_{\rm rad} +\rho_{\rm b} + \frac{1}{2} \left\{\left( {\displaystyle \frac{d \sigma}{dt}} \right)^2 + \omega^2 \sigma^2 + m^2 \sigma^2\right\} + \alpha \exp\left(-\frac{1}{2}\beta \sigma^2\right)  \right] \;\; ,
\end{equation}
where $\rho_{rad}$ is the usual density of relativistic photons and neutrinos, $\rho_b$ is the usual baryonic density, and $\omega = d\theta / dt$ is time-dependant. The second equation implies the conservation of the charge per comoving volume $Q = \omega \sigma^2 a^3$. Therefore, we can rewrite the first equation as
\begin{equation}
{\displaystyle \frac{d^2 \sigma}{d t^2}} + \frac{3}{a} {\displaystyle \frac{da}{dt}}{\displaystyle \frac{d \sigma}{dt}} + m^2 \sigma - \alpha \beta \sigma \exp\left(-\frac{1}{2}\beta \sigma^2\right) - {\displaystyle \frac{Q^2}{\sigma^3 a^6}} = 0 \;\;.
\end{equation}
Using the same parameters as defined before, we can solve the equations and determine the evolution of the cosmological parameters $\Omega = \rho/\rho_0^c$ in function of the scale factor $a$. The initial value of the field density can be chosen arbitrarily and leads to an identical global behavior. The charge $Q$ has been fixed so that the equality scalar field density/radiation density occurs approximately at the same time as the equality dark matter/radiation in the standard model of cosmology. The cosmological evolution of the density of the scalar field in this context is plotted on figure \ref{cosmo_evolution}.\\
One can remark that the field follows a four--step evolution, whose first three steps had already been observed in \cite{arbey_cosmo}, and which is directly related to the form of the Klein-Gordon and Friedman equations:
\begin{itemize}
\item The first behavior, at very high redshift, occurs when the kinetic term of the scalar field dominates, leading to a $a^{-6}$ decay of the scalar field density. At this time, the scalar field density is very high, and the potential has a negligible influence on the evolution of the field. This step is in fact determined by the high expansion rate of the Universe, so that only the first two terms of equation (\ref{KGcosmo1}) dominates. As at the BBN time the field density is negligible in comparison to the radiation density, this behavior is in agreement with the BBN constraints.
\item This behavior stops as the expansion rate decreases, {\it i.e.} as the influence of the potential increases. A transition will occur when the potential will exactly compensate the kinetic terms in equation (\ref{KGcosmo1}). The scalar field density will then reach a plateau, leading to a cosmological constant behavior.
\item This cosmological constant behavior will continue until $H \sim m$. At this time, the mass term becomes important in the potential, and the field oscillates quickly in time. Its dynamics is then comparable to that of an effective real scalar field oscillating rapidly, and its time-averaged pressure vanishes. Thus, the field then behaves like pressureless matter. This transition occurs before matter-radiation equality.
\item Recently, at a late time in structure formation, as the densities continue to decrease and the exponential part of the potential becomes non-negligible, the oscillations have damped sufficiently so that the field can settle into its non-degenerate vacuum. This leads to a dark energy behavior at low redshifts. However, the transition to this dark energy behavior is not completed yet, and in the future, one can expect the scalar field to achieve its transition and behave like a cosmological constant.
\end{itemize}
This scenario is in complete agreement with the constraints on a unifying dark fluid model which were derived in \cite{arbey_darkfluid}.

\section{Conclusion}
\noindent In this work, we have shown that it is possible to elaborate a dark fluid model replacing the standard dark energy / dark matter models, using a complex scalar field. We have seen that a potential containing a quadratic term and a Gaussian term is suitable to explain the gravitation observations at galaxy, cluster and cosmological scales. Based on \cite{arbey_quadratic}, we also know that this model can account for a large variety of rotation curve shapes. Thus, we can conclude that this model could be an interesting alternative to the dark matter / dark energy models.\\
Of course, many studies are still needed to test this model. In particular, it would be interesting to consider the formation of large scale structures and to study the influence of the dark fluid on the anisotropies of the Cosmic Microwave Background. Another question concerns the form of the potential. Indeed, this potential has interesting properties, but is not directly derived from high energy theories. We can consider it as a toy-potential rather than a definitive answer. It would then be interesting to perform the study of different kinds of potentials, and perhaps to find an adequate potential derived from a high energy theory.\\
In any case, this simple model looks promissing, and reinforces the suggestion that dark fluid models should be studied deeply.

\section*{Acknowledgements}
\noindent I would like to thank Alan Coley, Farvah Mahmoudi and Patrick Peter for useful discussions and comments.


\end{document}